\documentclass[notitlepage,letterpaper]{article}
\usepackage[ansinew]{inputenc} 
\usepackage{graphicx}
\usepackage{geometry}      
\usepackage{epstopdf}
\begin{document}

\title{\textbf{e-Science initiatives in Venezuela}}
\author{
\textbf{J. L. Chaves, G. Díaz, V. Hamar, R. Isea,} \\\textbf{ F. Rojas, N. Ruíz, R. Torrens, M. Uzcátegui }\\
Centro Nacional de Cálculo Científico Universidad de Los Andes \textsc{(CeCalCULA)}, \\
Corporación Parque Tecnológico de Mérida, Mérida 5101, Venezuela \\
\textbf{J. Flórez-López} \\
Departamento de Ingeniería Estructural, Facultad de Ingeniería,\\
Universidad de Los Andes, Mérida 5101, Venezuela\\
\textbf{H. Hoeger}\\
Centro de Simulación y Modelado (CeSiMo), Facultad de Ingeniería, \\
Universidad de Los Andes, Mérida 5101, Venezuela, and \\
Centro Nacional de Cálculo Científico Universidad de Los Andes \textsc{(CeCalCULA)}, \\
Corporación Parque Tecnológico de Mérida, Mérida 5101, Venezuela\\
\textbf{C. Mendoza } \\
Centro de Física, Instituto Venezolano de Investigaciones Científicas (IVIC), \\
PO Box 21827, Caracas, 1020A, Venezuela, and \\
Centro Nacional de Cálculo Científico Universidad de Los Andes \textsc{(CeCalCULA)}, \\
Corporación Parque Tecnológico de Mérida, Mérida 5101, Venezuela\\
\textbf{L. A. Núñez} \\
Centro de Física Fundamental, Departamento de Física, Facultad de Ciencias, \\
Universidad de Los Andes, Mérida 5101, Venezuela, and \\
Centro Nacional de Cálculo Científico Universidad de Los Andes \textsc{(CeCalCULA)},  \\
Corporación Parque Tecnológico de Mérida, Mérida 5101, Venezuela \\
}
\date{Versión $\alpha 2.00$  January 2006}
\maketitle


\begin{abstract}
Within the context of the nascent e-Science infrastructure in Venezuela, we describe several web-based
scientific applications developed at the Centro Nacional de Cálculo Científico Universidad de
Los Andes \textsc{(CeCalCULA)}, Mérida, and at the Instituto Venezolano de Investigaciones
Científicas (IVIC), Caracas. The different strategies that have been followed for implementing
quantum chemistry and atomic physics applications are presented. We also briefly discuss a
damage portal based on dynamic, nonlinear, finite elements of lumped damage mechanics and a
biomedical portal developed within the framework of the \textit{E-Infrastructure shared between
Europe and Latin America} (EELA) initiative for searching common sequences and inferring their
functions in parasitic diseases such as leishmaniasis, chagas and malaria.
\end{abstract}


\section{Introduction}
The term ``e-Science'' was introduced by John Taylor in 2000 envisioning the new trends that were starting
to occur in global collaborations in key areas of science. It defines a set of computational (hardware
\& middleware) and data services that enable service oriented science
\cite{HeyTrefetthen2003,HeyTrefetthen2005,Foster2005}.
These infrastructures and facilities have made it possible to develop computational ``collaboratories"
\cite{KouzesMyersWulf1996},
defined  as  places where  scientists  work  together  to  solve  complex
interdisciplinary  problems despite  geographic  and  organizational  boundaries.  Such
collaboratories provide  uniform access  to  computational  resources,  services
and/or  applications.  They also expand the resources available to researchers, foment multidisciplinary
collaborations and problem solving, increase the efficiency of research and accelerate the dissemination of knowledge.

The IT hardware infrastructure to support these multidisciplinary and distributed collaborations include
high-speed networks, supercomputers, workstation clusters and new expensive shared
experimental/simulations facilities such as sensors, satellites, high-performance-computer simulations and
high-throughput devices, among others. The software environments allow a user to authenticate, submit a job,
monitor running jobs, manage input/output data through distributed file systems and visualize results.
The new computing environments and tools should support all these requirements, and must be presented to the scientific communities in terms of the applications themselves rather than in the form of complex computing
protocols. The grid must be viewed as a seamless extension of the user computer facilities regarding
both job execution/monitoring and data access/management. The recent move of the grid community to a
service-oriented architecture and the proposal for an Open-Grid-Services Architecture (OGSA) based on commercially supported web-services technology is therefore of great significance \cite{FosterEtal2003}.

In this paper we mainly concentrate on the portal functionalities and the different strategies we have been followed for implementing web-based scientific applications that are required to make e-Science
a reality in our region. We briefly describe some of the web-based scientific applications developed at the Centro Nacional de Cálculo Científico Universidad de Los Andes (\textsc{CeCalCULA}\footnote{ \texttt{http://www.cecalc.ula.ve/}}) and at the Instituto Venezolano de Investigaciones Científicas (IVIC\footnote{ \texttt{http://www.ivic.ve/}}). \textsc{CeCalCULA} was established in 1997 as a joint effort between the Universidad de Los Andes, the Fondo
Nacional para la Ciencia y la Tecnología and the Corporación Parque Tecnológico de Mérida for the transfer of computer-intensive technology in science and engineering projects.
In the last decade this national center has also provided the local scientific research community with
consulting services, computing power and IT training. It is considered a main asset of the National Academic
Network of Research Centers and National Universities\footnote{ \texttt{http://www.reacciun2.edu.ve/view/reacciun.php}} and has contributed to generate a
favorable atmosphere for innovation which has been reviewed in recent studies of multilateral
organizations\footnote{ \texttt{http://www.pnud.org.ve/idhn\_2002/idhn\_2002.htm}}. The Computational Physics
and Computational Chemistry Laboratories of IVIC have been heavy users of high performance computing (HPC)
and software and database developers since the beginning of the 90s, and therefore there is much current
interest in the possibilities of the new e-Infrastructure.

The structure of the paper is as follows. In Section~\ref{CeCalCULAHandson} the national and regional leadership of \textsc{CeCalCULA} in
organizing hands-on workshops on IT, HPC and networking is summarized. In Section~\ref{StrategiesWEBScientificApp}
we discuss strategies for adapting and upgrading legacy scientific applications to a web-based grid
environment. The Damage Portal an e-Engineering application of lumped damage mechanics based
on dynamic, nonlinear finite elements is presented in Section~\ref{PortalofDamage}, followed in
Section~\ref{BiomedEELAApp} by the Blast2EELA Biomedical Portal implemented within the
\textit{E-Infrastructure shared between Europe and Latin America} (EELA) initiative. Conclusions and future
projects are outlined in Section~\ref{Conclusions}.


\section{Emphasis on hands-on training}
\label{CeCalCULAHandson}
In the past ten years \textsc{CeCalCULA} has organized a string of national and regional (Caribbean Basin and Andean countries)
workshops and schools aimed at high-level researchers and professionals. The Workshop on \textit{New Techniques
and Tools for Computational
Sciences}\footnote{ \texttt{http://www.cecalc.ula.ve/adiestramiento/eventos/TALLECIE/tallecie.html}} held in December 1996
was the first workshop on scientific computing in Venezuela. It attracted more than a hundred HPC users in several
disciplines, initiating an important and irreversible trend in the country as it became the cornerstone in the
identity of a young academic and research community that used the computer as a fundamental scientific tool.
Additionally it served as the launchpad for \textsc{CeCalCULA} as a national HPC center. This first successful meeting was
followed up two years later by the \textit{First Latin-American Workshop on Parallelism and High Performance
Computing}\footnote{ \texttt{http://www.cecalc.ula.ve/adiestramiento/eventos/ELPCAR/elpcar.html}} which convened the
Caribbean and Andean regions. The \textit{Second Latin-American Workshop on Parallelism and High Performance
Computing}\footnote{ \texttt{http://www.cecalc.ula.ve/adiestramiento/eventos/ELPCAR2/elpcar2.html}} in December 2001
had an emphasis on the emerging area of grid computing. The \textit{Latin-American School in
High Performance Computing on Linux Clusters}\footnote{ \texttt{http://www.cecalc.ula.ve/HPCLC/}} (October 2003),
motivated by the high demand of a similar workshop held in Trieste, Italy, the previous year, focused
on computer array technologies (clusters) and was eminently hands-on. The \textit{First Latin-American Grid
Workshop}\footnote{ \texttt{http://www.cecalc.ula.ve/lag2004/}}, November 2004, covered grid concepts in a
theoretical and practical way. The \textit{First Latin-American Workshop for Grid
Administrators}\footnote{ \texttt{http://www.cecalc.ula.ve/lagaw2005}}, November 2005, led to the launch of several
grid projects at national and Latin American levels, and was tailored for the technical personnel responsible
for grid infrastructure management.

January 2006 marked the official start of EELA\footnote{ \texttt{http://www.eu-eela.org/}} which will interconnect
Latin America to the European grids (EGEE), project of which the Universidad de Los Andes (ULA) is a partner. The \textit{Second
Latin-American Grid Workshop, First Latin American EELA Workshop and First Latin American EELA
Tutorial}\footnote{ \texttt{http://www.cecalc.ula.ve/lag2006/}} (April 2006) focused on the impact of grid
technologies on e-Science and on the advances in computational grids and their relation with different areas such as
data storage, computational visualization and distant collaborations. It also provided technical and practical
training and a space for discussion for EELA related issues. On July 2007, ULA will host the \textit{Second EELA
Grid School}\footnote{ \texttt{http://www.eu-eela.org/egris1/}}, a two-week hands-on activity in which participants
will work closely together with tutors in the grid porting of applications.

Another well known event hosted by ULA is the \textit{Latin-American Network
School}\footnote{ \texttt{http://www.eslared.org.ve/}} on its 9th edition since 1992 which provides participants
with up-to-date knowledge on networks, IT, security, open software, among others. ULA has had a
long tradition in networking, being one of the first Venezuelan universities to join Internet and has provided
training and support to many institutions in Venezuela and abroad. ULA also hosts the national laboratory on IPv6
and grid.

\section{Strategies for web-based scientific applications}
\label{StrategiesWEBScientificApp}
\begin{figure}[h]
\includegraphics[width=3in]{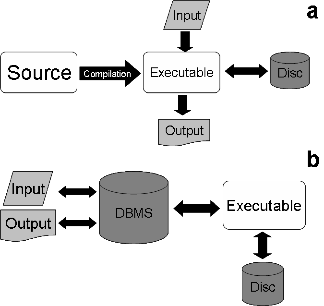}
   \caption{(a)Traditional scientific application. (b) Web-based scientific application (database centered).}
   \label{database}
\end{figure}

In spite of the spectacular evolution in computing capabilities brought about by
microcomputers, the Internet and the web in the past 15 years, scientific computing has not changed
much from the earlier days. As shown in Fig.~\ref{database}a, most legacy scientific applications are
monolithic fortran sources which are compiled locally; a usually complicated input file is then read at
running time to produce one or more disk files and a lengthy output file of numerical tables.
Input/output manipulation is usually performed with a text editor. Doing research with such computational
tools usually implies a long learning curve and much acquired expertise.
In the Computational Physics and Computational Chemistry Laboratories at IVIC, several
suites of codes are regularly used to carry out large calculations in atomic structure
(\textsc{autostructure} \cite{Badnell1997,EissnerJonesNussbaumer1974}),
electron impact scattering (\textsc{R-matrix} \cite{BerringtonEtal1978}) and quantum chemistry
(\textsc{cativic} \cite{RuetteEtal2004}, \textsc{gaussian-98} \cite{FrischEtal2002}). Considerable
effort has been recently dedicated to adapt these codes to the new grid environments,
namely developing portals, parallelization and code restructuring. In reference to
implementing web-based user interfaces, it was soon realized that the traditional computing paradigm
shown in Fig.~\ref{database}a was impractical and needed to evolve to a database-centered scheme
(Fig.~\ref{database}b). In the latter all the input/output manipulation and runtime job monitoring
is performed through a Database Management System (DBMS), e.g. MySQL, and thus submitting jobs
in HPC would not be much different from buying a book in \texttt{amazon.com} or placing a bid in
 \texttt{ebay.com}.

\begin{figure}[h]
\includegraphics[width=3in]{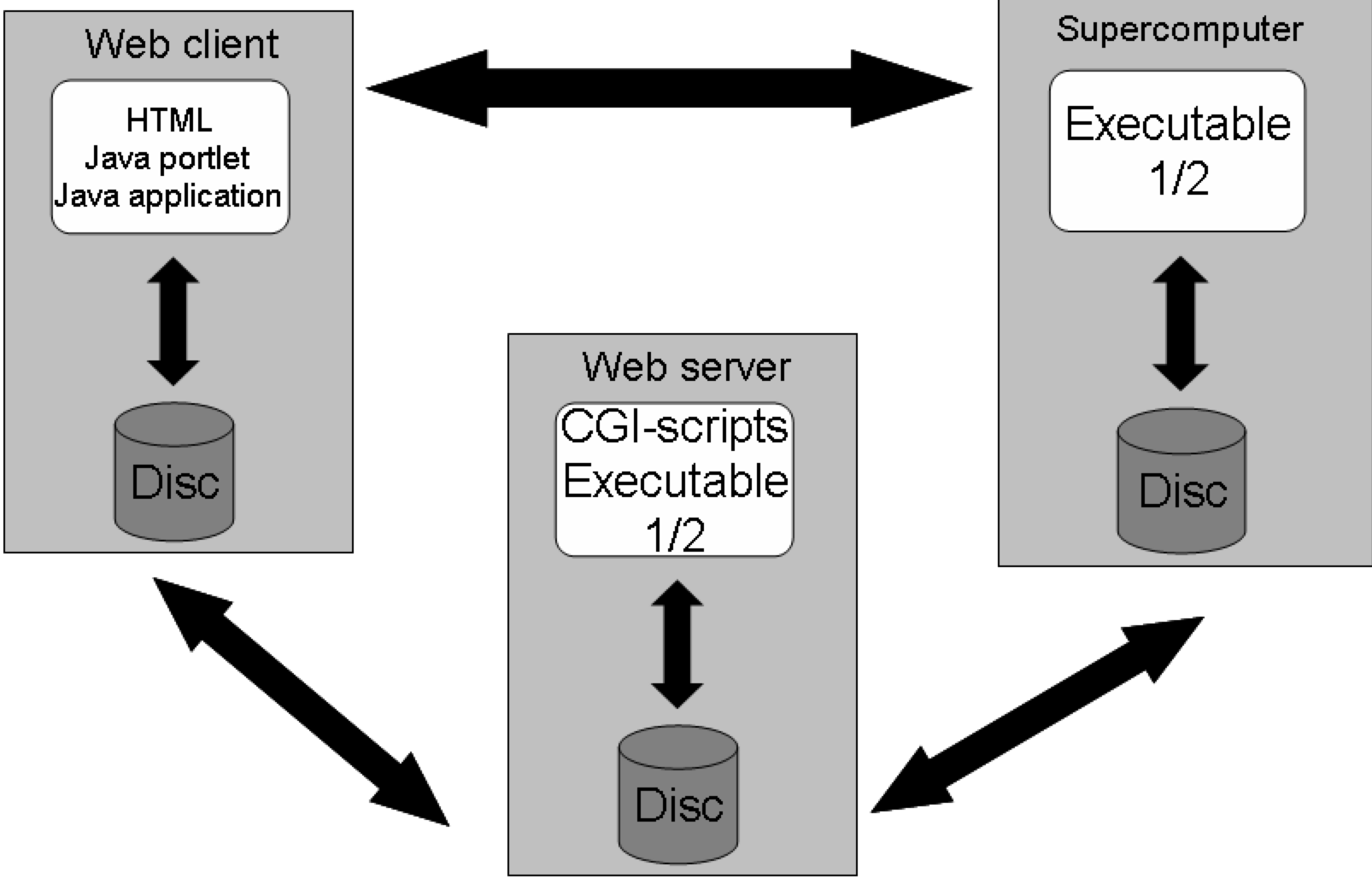}
   \caption{Distributed architecture of a web-based scientific application.}
   \label{arch}
\end{figure}

A second finding encountered in adapting scientific codes to the grid was that extensive code
restructuring and upgrading was unavoidable. The processor where the number crunching is carried out,
ideally a massively parallel cluster, is very different from the web server that houses web pages,
manages user interactivity and runs PHP or JSP scripts and also different from the user workstation where
a browser is loaded to run Javascripts and Java applets and applications. In some cases, even
the fortran routines had to be redistributed on the different processor types, and interface procedures developed
for network communication among them. The ideal new architecture is the triangular client--server model
shown in Fig.~\ref{arch}. Most portal functionalities have been coded with JSP, but in the case of
\textsc{cativic} that requires a molecular builder, a full Java application was developed.
Moreover, in all cases it was found that inter-processor communication must be reduced to URL
requests through port 80 in order to avoid site firewalls and port restrictions, and that
\textsc{cativic}'s Java application running at the user end communicates with both the web server
and the back-end supercomputer.
Alpha prototypes of the above listed codes, developed at IVIC by J. González, L.S. Rodríguez, M. Oldenhof and
G. Martorell, are currently operational and at the testing stage.

\section{Structural e-Engineering Portal of Damage}
\label{PortalofDamage}
\begin{center}
\begin{figure}[h]
\includegraphics[width=3in]{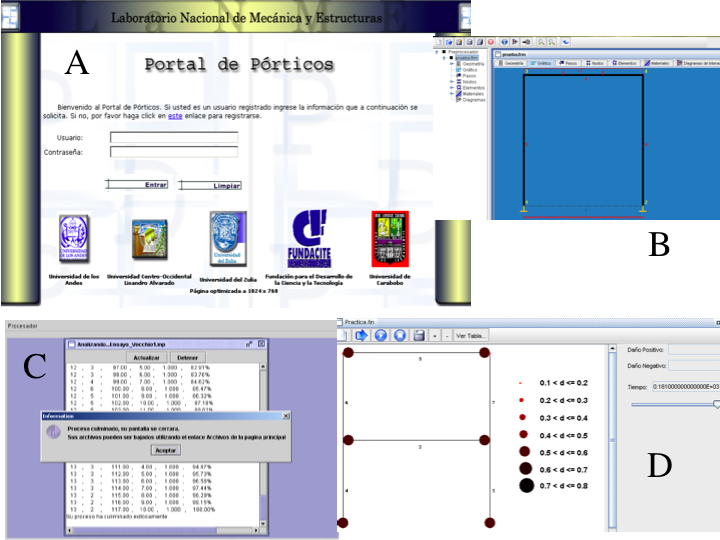}
   \caption{Portal of Damage. Plate A: Portal of Damage Homepage   \texttt{http://portaldeporticos.ula.ve/}. Plate B: Pre-processor. Plate C:  Monitoring of an analysis through the portal. Plate D: Graphic post-processor}
   \label{FigPortalDamage}
\end{figure}
\end{center}

The Damage Portal\footnote{ \texttt{http://portaldeporticos.ula.ve}} is a web-based finite element working environment for structural analysis described in detail elsewhere\cite{MaranteEtal2005} and depicted in Plate A, Fig.~\ref{FigPortalDamage}. It allows the user to numerically simulate cracking processes and collapses of reinforced concrete structures subjected to mechanical overloads, e.g. earthquake loadings. This system consists of a set of Java (working environment) and Fortran (generator engine) modules. The preprocessor Java module provides the environment for building the input structure and for evaluating its load (see Plate B, Fig.~\ref{FigPortalDamage}). This module generates a file containing the raw data and sends part of it to the generator (a piece of code that transforms these untreated data into information that can be used by the finite element simulator) to be refined. With these refined data, the preprocessor creates an input file for the analysis of the structure. All these input files can be downloaded by the user. Next, the refined data file becomes the input to the finite element simulator through a Java interface that allows the user to monitor, or to
abort, the analysis in Plate C, Fig.~\ref{FigPortalDamage}. The simulator is a dynamic, nonlinear finite element program written in Fortran, whose physical model is based on a new theory referred to as Lumped Damage Mechanics \cite{CipollinaLopezFlorez1995}. The simulator computes and quantifies the density and location of concrete cracking and reinforcement yielding a set of state variables. In particular, the concrete cracking density is described by a damage variable that can take values between zero (no damage) and one (complete concrete destruction).
The nonlinear dynamic analysis is carried out in a step-by-step procedure where the state of the structure
is determined during loading history. By examining the damage distribution, the user can determine the state
of reparability of the structure and the possibility of structural collapse. Numerically this collapse is
defined by the absence of a mathematical solution that complies simultaneously with the equilibrium equations
and the constitutive laws that describe the material behavior of the reinforced concrete structure. The
results of the simulator are stored in the server in the form of text and postprocessing files that can
also be downloaded. The postprocessing files are used by the fourth element of the Portal, the Postprocessor:
a Java module that generates the visualization of the damage through distribution maps,
variable vs. variable and variable vs. time curves (see Plate D, Fig.~\ref{FigPortalDamage}). Additionally the Portal includes a tutorial, a user manual and theory write-ups. None of the programs in the systems is actually
downloaded by the user. The Portal has been successfully employed for the evaluation of existing structures \cite{TorresGonzalesMujica2007}  and construction codes \cite{Moreno2005}.

\section{Blast2EELA Biomedical Portal}
\label{BiomedEELAApp}
The functionality study of the different genes and regions is one of the most important efforts on
genome analyses. If the queries and the alignments are well designed, both functional and evolutionary
information can be inferred from sequence alignments since they provide a powerful way to compare novel sequences
with previously characterized genes.
\begin{center}
\begin{figure}[h]
\includegraphics[width=3in]{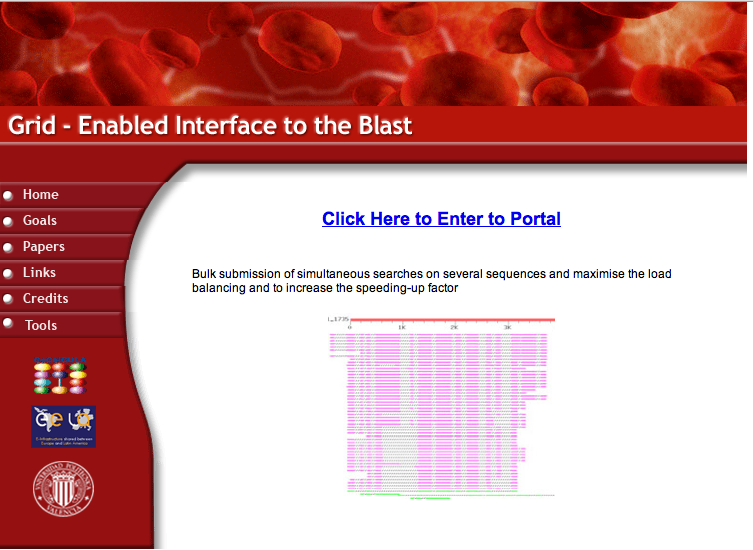}
   \caption{Blast2EELA Biomedical Portal  \texttt{http://www.cecalc.ula.ve/blast}}
   \label{BlastPortal}
\end{figure}
\end{center}
The Basic Local Alignment Search Tool (BLAST\footnote{ \texttt{http://www.ncbi.nlm.nih.gov/Education/BLASTinfo/information3.html}}) finds regions of local similarity between sequences. The program compares nucleotide or protein sequences against databases and calculates the statistical significance of the matches. This process of finding homologous sequences is computationally intensive since aligning a single sequence is not a costly task, but normally, thousands of sequences are searched simultaneously.

The biocomputing community usually relies on either local installations or public servers such as the
NCBI\footnote{ \texttt{http://www.ncbi.nlm.nih.gov/}} or the gPS@\footnote{ \texttt{http://gpsa.ibcp.fr/}}, but
the limitations on the number of simultaneous queries makes this environment inefficient for large tests.
Moreover, since the databases are updated periodically, it is convenient re-check the results of
previous studies. For this reason, we are developing within the EELA project \cite{CardenasEtal2006} a portal called Blast2EELA\footnote{ \texttt{http://www.cecalc.ula.ve/blast}} shown in Fig.~\ref{BlastPortal}. Through
this portal it is possible to have bulk submission of simultaneous searches on several sequences and to
improve its computational efficiency with the help of mpiBlast (i.e. a freely available open source
parallelization of NCBI Basic Local Alignment Search Tool \cite{KimEtal2005}).

The main input data for the Portal are only the sequences in Fasta format. It subsequently sends
the data to the grid and then displays the obtained results on the web interface for its interpretation. Every user has a private virtual work area, and therefore the Portal keeps the confidentiality of the data stored and sent through the grid. The user can customize the virtual work area accessible through a login and password. Once the user is authenticated, the blast portal issues a proxy with the user's digital credentials (X509 certificates) by means of the Grid Security Infrastructure (GSI) libraries, avoiding successive validation during the lifetime of the proxy. This portal has shown to be very easy to use without increasing the complexity of the site. BLAST in Grid (BiG) has been used for searching similar sequences and inferring their function in parasite diseases such as Leishmaniasis (mainly \textit{Mexican Leishmania}), Chagas (mainly \textit{Trypanosoma Cruzi}) and Malaria (mainly \textit{Plasmodium vivax}).


\section{Conclusions}
\label{Conclusions}
We have briefly described some of the efforts that \textsc{CeCalCULA} has dedicated to organize national
and regional workshops and schools aimed at high-level researchers and professionals in computational
science and engineering.  In the context of the new e-Infrastructure, we have also discussed some of the
web-based scientific applications developed at \textsc{CeCalCULA} and at IVIC. These pilot applications are
currently operational, and we keep encouraging users to move to web/grid environments. Therefore we will
continue to aggressively offer support for hands-on training initiatives, enroll user in grid experiences
and migrating their applications to provide an widely accessible infrastructure based on portals technologies and tools \cite{AndronicoEtal2003,ThomasBoisseau2003}   


\section*{Acknowledgments}
This work was partially funded by the European Commission through the EELA Project (E-Infrastructure shared
between Europe and Latin America). One of us (LAN), gratefully acknowledge the financial support of the
Fondo Nacional de Investigaciones Científicas y Tecnológicas under projects S1-2000000820 and
F-2002000426.

\bibliographystyle{plai}
\bibliography{EScienceSpain2006IEEEBib}

\end{document}